\documentclass[
amssymb,preprint,preprintnumbers,nofootinbib,superscriptaddress]{revtex4}
\usepackage{amsmath}
\usepackage{graphicx}
\usepackage{latexsym}
\usepackage{amsfonts}
\usepackage{url,hyperref}
\usepackage{bm}
\usepackage{textcomp}
\usepackage{xcolor}

\newcommand{\beq}{\begin{equation}}
\newcommand{\eeq}{\end{equation}}
\def\bea{\begin{eqnarray}}
\def\eea{\end{eqnarray}}

\begin{document}
\title{Quantum Brownian motion with non-Gaussian noises: Fluctuation-Dissipation Relation and nonlinear Langevin equation}
\author{Hing-Tong Cho}
\email[Email: ]{htcho@mail.tku.edu.tw}
\affiliation{Department of Physics, Tamkang University, Tamsui, Taipei, Taiwan}
\author{Bei-Lok Hu}
\email[Email: ]{blhu@umd.edu}
\affiliation{Joint Quantum Institute and Maryland Center for Fundamental Physics \\ University of Maryland, College Park, Maryland 20742-4111, USA}

\begin{abstract}
Building upon the work of Hu, Paz, and Zhang \cite{HPZ92,HPZ93} on open quantum systems we consider the quantum Brownian motion (QBM) model with one oscillator (position variable $x$) as the system,   {\it nonlinearly} coupled to an environment of $N$ harmonic oscillators (with mass $m_n$, natural frequency $\omega_n$, position  $q_n$ and  momentum $p_n$ variables)  in the form  $\sum_{n}\left(v_{n1}(x)q_{n}^{k}+v_{n2}(x)p_{n}^{l}\right)$ where $k, l$ are integers
(the present work only considers the $k=l=2$ cases). The vertex functions $v_{n1}, v_{n2} $ are of the form $v_{n1}=\lambda C_{n1} f(x), v_{n2}(x)=-\lambda\,C_{n2}m_{n}^{-2}\omega_{n}^{-2}f(x)$ where $C_{n1,2}$ are the coupling constants with the $n$th oscillator, $f(x)$ is any arbitrary function of $x$, and $\lambda$ is a dimensionless constant. Employing the closed-time-path formalism  the influence action  $S_{IF}$ is calculated using a perturbative expansion in  $\lambda$. It is possible to identify the terms in $S_{IF}$ quadratic or higher in $\Delta(s)\equiv f(x_{+}(s))-f(x_{-}(s))$ to constitute the noise kernel, while terms linear in $\Delta$ to that of the dissipation kernel. The non-Gaussian noise kernel  gives rise to non-zero three-point correlation function of the corresponding stochastic force. The pathway presented here should be useful for the exploration of \textit{non-Gaussian properties of systems nonlinearly coupled with their environments}; examples in early universe cosmology and in quantum optomechanics (QOM) are mentioned.  A modified fluctuation-dissipation relation (FDR) is also established, which ensures the consistency of the model and the accuracy of results even at higher perturbative orders. Another result of significance is the derivation of a nonlinear Langevin equation which is expected to be useful for many open quantum system applications.
\end{abstract}
\date{\today}
\maketitle
\tableofcontents

\section{Introduction}\label{sec:intro}

In this paper we explore the stochastic dynamics of a quantum open system, focusing on the nonGaussianity (nG) due to the existence of nonlinearity either a) in the system, such as an anharmonic oscillator, or b) in the environment, such as a nonlinear bath, or c) in their nonlinear coupling. Here we consider case c) with weak nonlinearity,   using functional perturbative methods (the other two cases are not studied here, but can in principle be treated with similar approach).  We consider the standard quantum Brownian motion (QBM) model, where the system is represented by a harmonic oscillator with configuration space variable $x$, nonlinearly coupled with an environment (bath) of $N$ harmonic oscillators with variables $q_n$, whose effect can be described by  {non-Markovian dissipation as well as} a set of nonGaussian noises acting on the system. For linear coupling of the type ${C_n x q_n}$ where $C_n$ is the coupling constant of the system with the $n$th bath oscillator, because the combined system + environment Hamiltonian is in a quadratic or Gaussian form, its dynamics can be solved completely, see, e.g.,\cite{HPZ92} for the derivation of an exact master equation for QBM. 

\vskip 10pt
\noindent{\it 1. Nonlinear coupling and perturbative treatment}

For weakly nonlinear couplings of the type $\lambda C_n f(x)q_n^k$, where $f(x)$ denotes any arbitrary function of $x$ and $k$ is an integer (polynomial in $q$) one can use the functional perturbative method designed in \cite{HPZ93} to derive the dynamics of the open system in successive orders of $\lambda$, a dimensionless constant measuring the strength of nonlinearity. In Ref \cite{HPZ93} the authors carried out calculations up to the second order in $\lambda$, presented the FDRs of colored multiplicative noises of different orders (Sec. III) and derived a master equation for the case $f(x)= x^2$ (Sec. IV.C).   Here, we go to the third order in $\lambda$ which enables us to derive three point functions of the nonGaussian noises and   the corresponding nonlinear Langevin equation    needed for exploring a range of problems involving nonlinearity and nonGaussianity.   

NonGaussianity in cosmology is an actively pursued topic since the seminal work of Maldacena \cite{Maldacena} who calculated the power spectrum of the three point correlation functions for primordial scalar and tensor fluctuations in single field inflationary models.  According to Weinberg \cite{Weinberg} that work amounted to the evaluation of tree graphs in the `in-in' (otherwise known as the Schwinger-Keldysh   or the closed time path (CTP) \cite{ctp}) formalism and he added on the consideration of loop graphs. These quantities should be  measurable in the spectrum of anisotropies of the cosmic microwave background. Calculation of correlation functions in stochastic inflation was presented in \cite{VenSta} using stochastic methods.  The three point functions our calculations can produce are of the same nature, except for the origin of nonGaussianity: the nonlinearity in these cosmological studies resides in the cubic term of the Lagrangian (from cubic scalars to cubic gravitons and combinations in between) of the system, corresponding to case a) we mentioned above whereas in our present study nonlinearity resides in the coupling between the systems and the environment, corresponding to case c) above.

Another difference from most works on   cosmological nonGaussianity so far is that the authors usually work in a closed system setting, with no concern of the environment their system interacts with, albeit there is self-interaction in the system (e.g., scalar field with gravitons). However, if one wants to find out the influence of another field present on the behavior of the inflaton field bearing observational consequence, the open system setting presented here, together with the CTP and the Feynman-Vernon influence functional (IF) \cite{if} approach would prove to be useful \footnote{For the treatment of backreaction effects of particle creation on the dynamics of the early universe using CTP or in-in methods, see \cite{CalHu87}. Implicit in it is regarding the quantum field as environment and the universe as the system. In fact a FDR between these two parties can be established in this semiclassical cosmology setting \cite{HuSin}. Open quantum systems approach in calculating the effects of stress tensor correlators (or quantum noise in a Gaussian setting \cite{if}) in quantum fields (as environment) on the dynamics of spacetime (as system) has been established in the  semiclassical stochastic gravity theory, see \cite{HuVerBook} and references therein.  QBM model and CTP-IF methods have been used for the treatment of decoherence of quantum fields, structure formation and entropy generation in cosmology, to name a few important processes.  For sample earlier and current works, see, e.g.,  \cite{HPZdec,LomMaz,RouVer} and \cite{HHentropy,Colas}.}.   
Examples of how quantum fields affect the inflaton dynamics are given in \cite{WNLLC07,LNWW11,ChoNg20}

Beyond going to a higher perturbative order, in our calculations we have also included nonlinear coupling in the momentum of the schematic form {$ p_{n}^2 + q_{n}^2$ (see Eq.~(\ref{Sint1})}. Quantum Brownian motion of a charged particle in a magnetic field involves both coupling via position and momentum variables \cite{SSinha}. Adding a bilinear momentum coupling term to existing QBM models with bilinear configuration space coupling has been studied in \cite{HuaZha22}.  Here we go to higher nonlinear perturbative orders in both coordinate and momentum couplings. This is useful for studying nonlinearly coupled systems and nonGaussianity in different settings in many areas of current research. Below we shall describe another example, namely, in QOM. 

\vskip 10pt
\noindent{\it 2. Fluctuation-dissipation relations}

FDR in an open system  is a consistency condition. The relation between the Hadamard and the causal Green functions governing the noise and dissipation kernels in the environment is mathematically locked in. In physical terms it is a relation between the fluctuations or noises in the environment and the dissipation they induce  in the system dynamics. Demonstrating its existence is a self-consistency check. It also tells something important about  the system's nonequilibrium dynamics, namely, the FDR is  predicated upon an open quantum system reaching a stationary state, given sufficiently long time for it to interact with an environment.  For example, in the case of a system described by an anharmonic oscillator interacting linearly with an environment described by a scalar quantum field, the authors of Ref \cite{HHnlFDR} presented general nonperturbative expressions for the rate of energy (power) exchange between the anharmonic oscillator and its thermal bath. For the cases that a stable final equilibrium state exists, and the nonstationary components of the two-point functions of the anharmonic oscillator have negligible contributions to the power balance, these authors showed nonperturbatively that equilibration implies an FDR for the anharmonic oscillator.  FDRs for  nonlinear couplings of the form ${C_n f(x) q_n^k}$ have been derived in \cite{HPZ93}. We shall show in Sec. IV modified FDR  to third order in $\lambda$. For investigations of FDR in dynamical Casimir effect (DCE) in a similar vein, see \cite{But2M,ButND}.

\vskip 10pt
\noindent{\it 3. Nonlinear Langevin equation}  

This paper ends with the derivation of a nonlinear Langevin equation. We mention here briefly the background.  For classical nonlinear Langevin equations, see the classic paper of Zwanzig \cite{Zwanzig}. In the quantum realm, an example of it is a paper by Diosi \cite{Diosi}, where a simple quantum Langevin equation in the Ito form of a stochastic differential equation is presented for the study of quantum Brownian motion. The scope of application is vast, we shall just mention one example of where quantum nonlinear Langevin equation is applied to current research. In relativistic heavy ion collision (RHIC) physics  a  generalized nonlinear Langevin equation is used in \cite{JinHu} to study the approach to QCD critical point, where  nonlinear interactions of quantum fluctuations is expected to play a significant role. Another example is in the  study of QOM. 

\vskip 10pt
\noindent{\it 4. Quantum Optomechanics as a sample application} 

Optomechanics deals with the interaction between light  in a cavity and the  mirrors forming the cavity. For a static configuration one would see Casimir effect and if either mirror is allowed to move, the DCE. The coupling between light and the mirror is usually assumed to be through the  radiation pressure of photons impinging on the mirror, causing a displacement x  of the mirror \cite{Law}. Investigations of the effective dynamics of the two mirrors interacting with the radiation pressure of photons in a cavity  using the functional perturbative method is reported in \cite{But2M}.  With the radiation pressure being proportional to the number $N$ of photons, the interaction is in the  $N x$ form\footnote{In the present context $x$ is regarded as the mechanical or external degree of freedom (dof) of the mirror. In actuality it is the internal dof which interacts with a quantum field, like the electronic levels of an atom,  and one needs to translate how that affects the external or mechanical dof, or vice versa, resulting in an effectively nonlinear interaction giving rise to nonMarkovian dynamics. This is a nontrivial procedure \cite{BarCal}. The $Nx$ interaction is in this sense a semi-phenomenological description.  A more refined description including all three dynamical variables, namely, the idf, mdf and the field, is by way of the MOF model \cite{GBH,SLH,SLS,ButND}}. If we represent the motion of the mirror by a harmonic oscillator and if we use a Fock space representation of the photon field in terms of the lowering $\hat b$ and raising operators $\hat b^\dagger$, then $\hat N= \hat b^\dagger\, \hat b$. Translating this to the canonical variables  {$p$ and $q$,
\begin{eqnarray}
\hat{b}=\frac{1}{\sqrt{2\hbar\omega}}(\omega q+ip)\ \ \ ;\ \ \
\hat{b}^{\dagger}=\frac{1}{\sqrt{2\hbar\omega}}(\omega q-ip).
\end{eqnarray}
The number operator
\begin{eqnarray}
    \hat{N}=\hat{b}^{\dagger}\,\hat{b}=\frac{1}{2\hbar\omega}(\omega^{2}q^{2}+p^{2}+\hbar),
\end{eqnarray}
and} we see that  both $p^2$ and $q^2$ participate.  This is why we wanted to include the momentum coupling. Since the interaction term we are considering has $f(x)$ the results we present here includes higher order in the mirror's displacement. Second order correction to the mirror displacement under the radiation pressure interaction is considered in \cite{But25}. Since we aim at the third order in $\lambda$ many nonGaussian properties of QOM arising from nonlinear coupling between light and mirror displacement can in principle be captured also. For a recent discussion of nonGaussianity in QOM, see, e.g., \cite{nGQOM}

This paper is organized as follows.  {Section II details the derivation of the influence functional. Due to the nonlinear coupling terms in our model, we employ a perturbative approach, expanding the influence action in powers of the coupling constant. In Section III, we interpret the effects of the influence action as those of a non-Gaussian noise and analyze its properties. The remaining effects, characterized as a dissipative force, are presented in Section IV, where we also derive the modified FDR between the noise and dissipation kernels. Section V presents the derivation of the nonlinear Langevin equation and provides perturbative solutions under specific boundary conditions. Finally, we offer concluding remarks and further discussion in Section VI.} 

 
\section{Perturbative expansion of the influence functional}\label{sec:perturb}

To implement our idea in a concrete model, we consider the system of a Brownian particle with nonlinear coupling to the environment as elaborated in the paper by Hu, Paz, and Zhang \cite{HPZ93}. We shall largely follow their notation (with later modifications). In this model the particle action is given by
\begin{eqnarray}
S[x]=\int_{0}^{t}ds\left[\frac{1}{2}M\dot{x}^{2}-V(x)\right],
\end{eqnarray}
while the environment consists of a set of harmonic oscillators
\begin{eqnarray}
S_{e}[\{q_{n}\}]=\int_{0}^{t}ds\sum_{n}\left[\frac{1}{2}m_{n}\dot{q}_{n}^{2}-\frac{1}{2}m_{n}\omega_{n}^{2}q_{n}^{2}\right].
\end{eqnarray}
Note that for simplicity we shall take the temperature to be zero. The finite temperature case can be dealt with analogously. The interaction has the form
\begin{eqnarray}\label{Sint}
S_{int}[x,\{q_{n}\}]=\int_{0}^{t}ds\sum_{n}\left(v_{n1}(x)q_{n}^{k}+v_{n2}(x)p_{n}^{l}\right),
\end{eqnarray}
where $p_{n}$ is the momentum of the particle. In the following we shall take $k=l=2$. Models with other values of $k$ and $l$ can be analyzed in a similar way. Note that in \cite{HPZ93} only the first interaction term is present. Here we have included the momentum interaction term for broader application of our model. We shall give more specific examples in the following sections.

We define the vertex functions 
\begin{eqnarray}
    v_{n1}(x)=-\lambda\, C_{n1}f(x)\ \ \ ;\ \ \  v_{n2}(x)=-\lambda\,C_{n2}m_{n}^{-2}\omega_{n}^{-2}f(x)
\end{eqnarray}
with the dimensionless coupling constant $\lambda$. Since $p_{n}=m_{n}\dot{q}_{n}$, it is more convenient to rewrite the interaction action as
\begin{eqnarray}\label{Sint1}
    S_{int}[x,\{q_{n}\}]=\int_{0}^{t}ds\,(-\lambda\,f(x))\sum_{n}\left(C_{n1}q_{n}^{2}+C_{n2}\,\omega_{n}^{-2}\,\dot{q}_{n}^{2}\right)
\end{eqnarray}

The influence functional associated with the environment can be calculated using the closed-time-path formalism.
\begin{eqnarray}\label{CTPform}
F[x_{+},x_{-}]&=&\int_{CTP}\prod_{n}Dq_{n+}Dq_{n-}\ \!e^{\frac{i}{\hbar}\left\{S_{e}[\{q_{n+}\}]+S_{int}[x_{+},\{q_{n+}\}]-S_{e}[\{q_{n-}\}]-S_{int}[x_{-},\{q_{n-}\}]\right\}}\nonumber\\
&=&e^{\frac{i}{\hbar}S_{IF}[x_{+},x_{-}]},
\end{eqnarray}
where $S_{IF}[x_{+},x_{-}]$ is the influence action. This influence action can be expanded perturbatively using Feynman diagrams with the coupling constant $\lambda$ as an expansion parameter,
\begin{eqnarray}\label{IFexp}
S_{IF}[x_{+},x_{-}]=\sum_{j=1,2,3,..}S_{IF}^{(j)}[x_{+},x_{-}].
\end{eqnarray}
In ref.~\cite{HPZ93}, without the momentum interaction, the first two terms in Eq.~(\ref{IFexp}) are evaluated. Here we shall try to evaluate these terms to higher orders.

To facilitate the perturbative expansion, we write
\begin{eqnarray}
    F[x_{+},x_{-}]=\prod_{n}F_{n}[x_{+},x_{-}],
\end{eqnarray}
where 
\begin{eqnarray}\label{inffun}
    F_{n}[x_{+},x_{-}]
    &=&\left.e^{\frac{i}{\hbar}\left\{S_{int}\left[x_{+},\left\{\frac{\hbar}{i}\frac{\delta}{\delta J_{+}}\right\}\right]-S_{int}\left[x_{-},\left\{\frac{\hbar}{i}\frac{\delta}{\delta J_{-}}\right\}\right]\right\}}\ F_{n}^{(1)}[J_{+},J_{-}]\right|_{J_{+}=J_{-}=0}
\end{eqnarray}
and $F_{n}^{(1)}[J_{+},J_{-}]$ is the influence functional with the system oscillators linearly coupled to sources $J$.
\begin{eqnarray}\label{inffunn}
    &&F_{n}^{(1)}[J_{+},J_{-}]\nonumber\\
    &=&\int_{CTP}\prod_{n}Dq_{n+}Dq_{n-}\ \!e^{\frac{i}{\hbar}\left\{S_{e}[\{q_{n+}\}]+\int_{0}^{t}ds\,J_{+}(s)q_{+}(s)-S_{e}[\{q_{n-}\}]-\int_{0}^{t}ds\,J_{-}(s)q_{n-}(s)\right\}}\nonumber\\
    &=&e^{\frac{i}{2\hbar^{2}}\int_{0}^{t}ds\int_{0}^{s}ds'\,\left[J_{+}(s)G_{n++}(s,s')J_{+}(s')-J_{+}(s)G_{n+-}(s,s')J_{-}(s')-J_{-}(s)G_{n-+}(s,s')J_{+}(s')+J_{-}(s)G_{n--}(s,s')J_{-}(s')\right]}\nonumber\\
\end{eqnarray}
where the Schwinger-Keldysh propagators are
\begin{eqnarray}\label{SKpro}
    G_{n++}(s,s')&=&\hbar\left[-\mu_{n}(s-s'){\rm sgn}(s-s')+i\nu_{n}(s-s')\right]\nonumber\\
    G_{n+-}(s,s')&=&\hbar\left[\,\mu_{n}(s-s'){\rm sgn}(s-s')+i\nu_{n}(s-s')\right]\nonumber\\
    G_{n-+}(s,s')&=&\hbar\left[-\mu_{n}(s-s'){\rm sgn}(s-s')+i\nu_{n}(s-s')\right]\nonumber\\
    G_{n--}(s,s')&=&\hbar\left[\,\mu_{n}(s-s'){\rm sgn}(s-s')+i\nu_{n}(s-s')\right]
\end{eqnarray}
with
\begin{eqnarray}
\mu_{n}(s-s')=-\frac{1}{2m_{n}\omega_{n}}\sin[\omega_{n}(s-s')]\ \ \ ;\ \ \ \nu_{n}(s-s')=\frac{1}{2m_{n}\omega_{n}}\cos[\omega_{n}(s-s')].
\end{eqnarray}
From the form of the influence functional in Eq.~(\ref{inffun}), we can develop the expansion
\begin{eqnarray}
    F_{n}[x_{+},x_{-}]&=&\Bigg\{1+\frac{i}{\hbar}\left[S_{int}\left[x_{+},\left\{\frac{\hbar}{i}\frac{\delta}{\delta J_{+}}\right\}\right]-S_{int}\left[x_{-},\left\{-\frac{\hbar}{i}\frac{\delta}{\delta J_{-}}\right\}\right]\right]\nonumber\\
    &&\ \ +\frac{1}{2}\left(\frac{i}{\hbar}\right)^{2}\left[S_{int}\left[x_{+},\left\{\frac{\hbar}{i}\frac{\delta}{\delta J_{+}}\right\}\right]-S_{int}\left[x_{-},\left\{-\frac{\hbar}{i}\frac{\delta}{\delta J_{-}}\right\}\right]\right]^{2}\nonumber\\
     &&\ \ +\frac{1}{3!}\left(\frac{i}{\hbar}\right)^{3}\left[S_{int}\left[x_{+},\left\{\frac{\hbar}{i}\frac{\delta}{\delta J_{+}}\right\}\right]-S_{int}\left[x_{-},\left\{-\frac{\hbar}{i}\frac{\delta}{\delta J_{-}}\right\}\right]\right]^{3}\nonumber\\
    &&\left.\hskip 50pt +\cdots\Bigg\}F_{n}^{(1)}[J_{+},J_{-}]\right|_{J_{+}=J_{-}=0}
\end{eqnarray}
Therefore, the influence action in each perturbative order can now be expressed as
\begin{eqnarray}\label{PerExp}
    &&S_{IF}^{(j)}[x_{+},x_{-}]\nonumber\\
    &=&\sum_{n}\left.\frac{1}{j!}\left(\frac{i}{\hbar}\right)^{j}\Bigg\{S_{int}\left[x_{+},\left\{\frac{\hbar}{i}\frac{\delta}{\delta J_{+}}\right\}\right]-S_{int}\left[x_{-},\left\{-\frac{\hbar}{i}\frac{\delta}{\delta J_{-}}\right\}\right]\Bigg\}^{j}\ F_{n}^{(1)}[J_{+},J_{-}]\right|_{J_{+}=J_{-}=0,{\rm conn}}\nonumber\\
\end{eqnarray}
Note that the influence action is up in the exponential in Eq.~(\ref{CTPform}). Hence, we only considered the connected terms on the right-hand-side of the equation.

The first term $S_{IF}^{(1)}$ can be obtained by taking the functional derivatives above as in Eq.(\ref{PerExp}) with respect to $J_{+}$ and $J_{-}$, together with the expressions for the Schwinger-Keldysh propagators given in Eq.~(\ref{SKpro}).
\begin{eqnarray}
S_{IF}^{(1)}=\int_{0}^{t}ds\left(-\sum_{n}\delta V_{n1}^{(1)}(x_{+})\right)-\int_{0}^{t}ds\left(-\sum_{n}\delta V_{n2}^{(1)}(x_{-})\right),
\end{eqnarray}
where
\begin{eqnarray}
\delta V_{n1}^{(1)}(x_{+})&=&\lambda\left(\frac{\hbar}{2m_{n}\omega_{n}}\right)\left[2i\delta(0)\left(\frac{C_{n2}}{\omega_{n}}\right)+(C_{n1}+C_{n2})\right]f(x_{+}),\\
\delta V_{n2}^{(1)}(x_{-})&=&\lambda\left(\frac{\hbar}{2m_{n}\omega_{n}}\right)\left[-2i\delta(0)\left(\frac{C_{n2}}{\omega_{n}}\right)+(C_{n1}+C_{n2})\right]f(x_{-}).
\end{eqnarray}
They can be interpreted as renormalization terms of the potential in the order $\lambda$. As $\delta(0)$ is divergent, we have both infinite and finite renormalizations. It is interesting to note that the infinite renormalization is due to the momentum interaction term with the coefficient $C_{n2}$ and it is not present in \cite{HPZ93}.

The second term reads
\begin{eqnarray}\label{deltaa1}
S_{IF}^{(2)}&=&\int_{0}^{t}ds\left(-\sum_{n}\delta V_{n1}^{(2)}(x_{+})\right)-\int_{0}^{t}ds\left(-\sum_{n}\delta V_{n2}^{(2)}(x_{-})\right)\nonumber\\
&&\ \ -2\int_{0}^{t}ds\int_{0}^{s}ds'\ \!\Delta(s)\Sigma(s')\mu(s-s')+\frac{i}{2}\int_{0}^{t}ds\int_{0}^{t}ds'\ \!\Delta(s)\Delta(s')\nu(s-s'),
\end{eqnarray}
where 
\begin{eqnarray}
    \delta V_{n1}^{(2)}(x_{+})&=&\lambda^{2}\left(\frac{\hbar \,C_{n2}^{2}}{m_{n}^{2}\omega_{n}^{3}}\right)\left[i\delta(0)\left(\frac{1}{\omega_{n}}\right)+1\right]f^{2}(x_{+}),\\
    \delta V_{n2}^{(2)}(x_{-})&=&\lambda^{2}\left(\frac{\hbar\,C_{n2}^{2}}{m_{n}^{2}\omega_{n}^{3}}\right)\left[-i\delta(0)\left(\frac{1}{\omega_{n}}\right)+1\right]f^{2}(x_{-})
\end{eqnarray}
are the potential renormalization terms at order $\lambda^{2}$. In Eq.~(\ref{deltaa1}), we have also used the notation $\Delta(s)\equiv f(x_{+}(s))-f(x_{-}(s))$ and $\Sigma(s)\equiv [f(x_{+}(s))+f(x_{-}(s))]/2$. Here  $\nu(s-s')$ is the noise kernel,
\begin{eqnarray}\label{nu}
\nu(s-s')=\lambda^{2}\sum_{n}2\hbar\, (C_{n1}-C_{n2})^{2}\left[\nu_{n}^{2}(s-s')-\mu_{n}^{2}(s-s')\right].
\end{eqnarray}

The other quantity is the dissipation kernel
\begin{eqnarray}
\mu(s-s')=\lambda^{2}\sum_{n}4\hbar\, (C_{n1}-C_{n2})^{2}\,\mu_{n}(s-s')\nu_{n}(s-s'),
\end{eqnarray}
is related to the so-called damping kernel \cite{FJH}
\begin{eqnarray}
\gamma(s-s')=\lambda^{2}\sum_{n}\frac{\hbar}{\omega_{n}}(C_{n1}-C_{n2})^{2}
\left[\nu_{n}^{2}(s-s')-\mu_{n}^{2}(s-s')\right],\label{diszero}
\end{eqnarray}
with
\begin{eqnarray}
\mu(s-s')=\frac{d}{ds}\gamma(s-s').\label{mugamma}
\end{eqnarray}

In this work we go to higher orders of the perturbative expansion in $\lambda$ to explore the effect of the whole influence action. Using the corresponding Schwinger-Keldysh propagators we obtain
\begin{eqnarray}
S_{IF}^{(3)}
&=&\int_{0}^{t}ds\left(-\sum_{n}\delta V_{n1}^{(3)}(x_{+})\right)-\int_{0}^{t}ds\left(-\sum_{n}\delta V_{n2}^{(3)}(x_{-})\right)\nonumber\\
&&+\int_{0}^{t}ds\int_{0}^{s}ds'\ \Delta(s)\left(2\Sigma(s)\Sigma(s')+\Sigma^{2}(s')\right)\times\nonumber\\
&&\ \ \ \ \ \sum_{n}\lambda^{3}\left(\frac{16\hbar C_{n2}^{2}}{m_{n}\omega_{n}^{2}}\right)\left(C_{n1}-C_{n2}\right)\mu_{n}(s-s')\nu_{n}(s'-s)\nonumber\\
&&+\int_{0}^{t}ds\int_{0}^{s}ds'\int_{0}^{s'}ds''\
\Delta(s)\Sigma(s')\Sigma(s'')\sum_{n}\lambda^{3}\,(32\hbar)\,(C_{n1}+C_{n2})(C_{n1}-C_{n2})^{2}\times\nonumber\\
&&\hskip 50pt
\mu_{n}(s-s')\left[\nu_{n}(s'-s'')\mu_{n}(s''-s)
-\mu_{n}(s'-s'')\nu_{n}(s''-s)\right]\nonumber\\
&&+i\int_{0}^{t}ds\int_{0}^{s}ds'\ \Delta(s)\Delta(s')\left(\Sigma(s)+\Sigma(s')\right)\times\nonumber\\
&&\ \ \ \ \ \sum_{n}\lambda^{3}\left(\frac{8\hbar C_{n2}^{2}}{m_{n}\omega_{n}^{2}}\right)\left(C_{n1}-C_{n2}\right)\left(\mu_{n}^{2}(s-s')-\nu_{n}^{2}(s'-s)\right)\nonumber\\
&&+i\int_{0}^{t}ds\int_{0}^{s}ds'\int_{0}^{s'}ds''\
\Delta(s)\Sigma(s')\Delta(s'')\sum_{n}\lambda^{3}\,(16\hbar)\left(C_{n1}+C_{n2}\right)\left(C_{n1}-C_{n2}\right)^{2}\times\nonumber\\
&&\hskip 50pt
\mu_{n}(s-s')\left[\nu_{n}(s'-s'')\nu_{n}(s''-s)
+\mu_{n}(s'-s'')\mu_{n}(s''-s)\right]\nonumber\\
&&-i\int_{0}^{t}ds\int_{0}^{s}ds'\int_{0}^{s'}ds''\
\Delta(s)\Delta(s')\Sigma(s'')\sum_{n}\lambda^{3}\,(16\hbar)\left(C_{n1}+C_{n2}\right)\left(C_{n1}-C_{n2}\right)^{2}\times\nonumber\\
&&\hskip 50pt
\nu_{n}(s-s')\left[\nu_{n}(s'-s'')\mu_{n}(s''-s)
-\mu_{n}(s'-s'')\nu_{n}(s''-s)\right]\nonumber\\
&&+\int_{0}^{t}ds\int_{0}^{s}ds'\ \Delta(s)\Delta^{2}(s')\times\nonumber\\
&&\ \ \ \ \ \sum_{n}\lambda^{3}\left(\frac{4\hbar C_{n2}^{2}}{m_{n}\omega_{n}^{2}}\right)\left(C_{n1}-C_{n2}\right)\mu_{n}(s-s')\nu_{n}(s'-s)\nonumber\\
&&+\int_{0}^{t}ds\int_{0}^{s}ds'\int_{0}^{s'}ds''\
\Delta(s)\Delta(s')\Delta(s'')\sum_{n}\lambda^{3}\,(8\hbar)\,\left(C_{n1}+C_{2}\right)\left(C_{n1}-C_{n2}\right)^{2}\times\nonumber\\
&&\hskip 50pt
\nu_{n}(s-s')\left[\nu_{n}(s'-s'')\nu_{n}(s''-s)
+\mu_{n}(s'-s'')\mu_{n}(s''-s)\right].\label{deltaa2}
\end{eqnarray}
where 
\begin{eqnarray}
    \delta V_{n1}^{(3)}(x_{+})&=&\lambda^{3}\left(\frac{2\hbar C_{n2}^{3}}{m_{n}^{3}\omega_{n}^{5}}\right)\left[i\delta(0)\left(\frac{2}{3\,\omega_{n}}\right)+1\right]f^{3}(x_{+}),\\
    \delta V_{n2}^{(3)}(x_{-})&=&\lambda^{3}\left(\frac{2\hbar C_{n2}^{3}}{m_{n}^{3}\omega_{n}^{5}}\right)\left[-i\delta(0)\left(\frac{2}{3\,\omega_{n}}\right)+1\right]f^{3}(x_{-})
\end{eqnarray}

From Eqs.~(\ref{deltaa1}) to (\ref{deltaa2}) one can see the pattern in this expansion. To the next order in $\lambda$, one would have terms with $\Delta\Sigma\Sigma\Sigma$, $\Delta\Delta\Sigma\Sigma$, $\Delta\Delta\Delta\Sigma$, and $\Delta\Delta\Delta\Delta$. All the higher order terms can be obtained with the help of Feynman diagrams. However, we shall stop here and try to elucidate our idea on the influence action using the terms up to this order.

For terms with one factor of $\Delta$ as can be seen in Eqs.~(\ref{deltaa1}) to (\ref{deltaa2}), they can be grouped together as part of a generalized dissipation kernel. While terms with higher powers of $\Delta$ can be considered as part of a generalized noise kernel. In this respect, one could write the effective action which includes the Brownian particle action as well as the influence action as follows. 
\begin{eqnarray}\label{effact}
\Gamma[x_{+},x_{-}]&=&S[x_{+}]-S[x_{-}]-\int_{0}^{t}ds\,\delta V(x_{+})+\int_{0}^{t}ds\,\delta V(x_{-})\nonumber\\
&&\ \ -\int_{0}^{t}ds\,\Delta(s)J(s;\Sigma)
+\frac{i}{2}\int_{0}^{t}ds\int_{0}^{t}ds'\,\Delta(s)\Delta(s')N_{2}(s,s';\Sigma)\nonumber\\
&&\ \ -\frac{1}{3!}\int_{0}^{t}ds\int_{0}^{t}ds'\int_{0}^{t}ds''\,\Delta(s)\Delta(s')\Delta(s'')N_{3}(s,s',s'';\Sigma)+\cdots,
\end{eqnarray}
where 
\begin{eqnarray}
    \delta V_{1}(x_{+})&=&\sum_{n}\delta V_{n1}^{(1)}(x_{+})+\sum_{n}\delta V_{n1}^{(2)}(x_{+})+\sum_{n}\delta V_{n1}^{(3)}(x_{+})+\cdots\\
    \delta V_{2}(x_{-})&=&\sum_{n}\delta V_{n2}^{(1)}(x_{-})+\sum_{n}\delta V_{n2}^{(2)}(x_{-})+\sum_{n}\delta V_{n2}^{(3)}(x_{-})+\cdots
\end{eqnarray}
and the kernels
\begin{eqnarray}
J(s;\Sigma)&=&J^{(0)}(s;\Sigma)+J^{(1)}(s;\Sigma)+\cdots\nonumber\\
&=&\int_{0}^{t}ds'\,J^{(0)}(s,s')\,\Sigma(s')+\int_{0}^{t}ds'\int_{0}^{t}ds''\,J^{(1)}(s,s',s'')\,\Sigma(s')\,\Sigma(s'')+\cdots\nonumber\\  \label{Jkernel}\\
N_{2}(s,s';\Sigma)&=&N_{2}^{(0)}(s,s')+N_{2}^{(1)}(s,s';\Sigma)+\cdots\nonumber\\
&=&N^{(0)}_{2}(s,s')+\int_{0}^{t}ds''\,N^{(1)}_{2}(s,s',s'')\,\Sigma(s'')+\cdots\\
N_{3}(s,s',s'';\Sigma)&=&N_{3}^{(1)}(s,s',s'')+\cdots,\label{kernels}
\end{eqnarray}
with
\begin{eqnarray}
J^{(0)}(s,s')&=&\lambda^{2}\,\sum_{n}8\hbar\,(C_{n1}-C_{n2})^{2}\,\theta(s-s')\,\mu_{n}(s-s')\,\nu_{n}(s-s')\nonumber\\
&=&2\,\theta(s-s')\,\mu(s-s'),\nonumber\\
J^{(1)}(s,s',s'')&=&-\lambda^{3}\,\sum_{n}8\hbar\,(C_{n1}-C_{n2})\nonumber\\
&&\bigg\{\theta(s-s')\,\mu_{n}(s-s')\Big[\frac{C_{n2}^{2}}{m_{n}\omega_{n}^{2}}\big(2\delta(s-s'')+\delta(s'-s'')\big)\nu_{n}(s'-s)\nonumber\\
&&\ \ \ \ \ \ +2(C_{n1}^{2}-C_{n2}^{2})\,\theta(s'-s'')\Big(\nu_{n}(s'-s'')\,\mu_{n}(s''-s)\nonumber\\
&&\hskip 180pt
-\mu_{n}(s'-s'')\,\nu_{n}(s''-s)\Big)\Big]\nonumber\\
&&\hskip 50pt 
+(s'\leftrightarrow s'')\bigg\}, \label{Jkernels}
\\
N^{(0)}_{2}(s,s')&=&\lambda^{2}\,\sum_{n}2\hbar\,(C_{n1}-C_{n2})^{2}\left[\nu_{n}^{2}(s-s')-\mu_{n}^{2}(s-s')\right]\nonumber\\
&=&\nu(s-s'),\nonumber\\
N^{(1)}_{2}(s,s',s'')&=&\lambda^{3}\sum_{n}8\hbar\,(C_{n1}-C_{n2})\nonumber\\
&&\bigg\{\Big[-\frac{C_{n2}^{2}}{m_{n}\omega_{n}^{2}}\,\delta(s-s'')\big(\nu_{n}^{2}(s-s')-\mu_{n}^{2}(s-s')\big)\nonumber\\
&&\ \ \ \ \ \ +2(C_{n1}^{2}-C_{n2}^{2})\,\theta(s-s'')\,\mu_{n}(s-s'')\nonumber\\
&&\hskip 50pt \Big(\nu_{n}(s''-s')\nu_{n}(s'-s)+\mu_{n}(s''-s')\mu_{n}(s'-s)\Big)\Big]\nonumber\\
&&\hskip 70pt+(s\leftrightarrow s')\bigg\},\label{noisen2}\\ 
N^{(1)}_{3}(s,s',s'')&=&-\lambda^{3}\,\sum_{n}4\hbar\,(C_{n1}-C_{n2})\nonumber\\
&&\bigg\{\theta(s-s')\,\nu_{n}(s-s')\Big[-\frac{C_{n2}^{2}}{m_{n}\omega^{2}_{n}}\,\delta(s'-s'')\,\mu_{n}(s''-s)\nonumber\\
&&\hskip 120pt +2(C_{n1}^{2}-C_{n2}^{2})\,\theta(s'-s'')\nonumber\\
&&\hskip 100pt \Big(\nu_{n}(s'-s'')\nu_{n}(s''-s)+\mu_{n}(s'-s'')\mu_{n}(s''-s)\Big)\Big]\nonumber\\
&&\hskip 50pt +(5\ {\rm permutations\ of\ } s,s',s'')\bigg\}.\label{kerter}
\end{eqnarray}

Eq.~(\ref{effact}) summarizes the effective action of the Brownian particle including the effects coming from its interaction with the set of harmonic oscillators in the environment. In the following sections we shall try to interpret these terms as dissipation and noise effects from the environment.

%
%
%

\section{Non-Gaussian stochastic forces}\label{sec:stochastic}

The terms quadratic or higher in powers of $\Delta(s)$ in Eq.~(\ref{effact}) can be interpreted as the effect of a single stochastic force $\xi(s)$. This is accomplished by writing
\begin{eqnarray}
&&e^{\frac{i}{\hbar}\left[\frac{i}{2}\int_{0}^{t}ds\int_{0}^{t}ds'\,\Delta(s)\Delta(s')N_{2}(s,s';\Sigma)
-\frac{1}{3!}\int_{0}^{t}ds\int_{0}^{t}ds'\int_{0}^{t}ds''\,\Delta(s)\Delta(s')\Delta(s'')N_{3}(s,s',s'')\right]}\nonumber\\
&=&\int D\xi\, P[\xi]\ e^{\frac{i}{\hbar}\int_{0}^{t}ds\,\Delta(s)\xi(s)},\label{stodef}
\end{eqnarray}
where $P[\xi]$ is the probability density of the stochastic force $\xi(s)$. 

Comparing the powers of $\Delta(s)$ on both sides of Eq.~(\ref{stodef}) it is possible to establish the stochastic average of various correlators of the $\xi(s)$. We shall consider the lowest few correlators. For one $\xi$,
\begin{eqnarray}
\left\langle\xi(s)\right\rangle=\int D\xi\, P[\xi]\,\xi(s)=0.
\end{eqnarray}
For two $\xi$'s,
\begin{eqnarray}
\left\langle\xi(s)\xi(s')\right\rangle=\int D\xi\, P[\xi]\,\xi(s)\xi(s')
=\hbar N_{2}(s,s';\Sigma).
\end{eqnarray}
From Eqs.~(\ref{kernels}) and (\ref{kerter}), one can see that to order $\lambda^{2}$ this correlator is given by
\begin{eqnarray}\label{n20}
\left\langle\xi(s)\xi(s')\right\rangle^{(0)}=\hbar\,N_{2}^{(0)}(s,s')
=\hbar\,\nu(s-s').
\end{eqnarray}
To all orders one should regard the full two-point correlation function $N_{2}(s,s';\Sigma)$ in Eq.~(\ref{kernels}) as the noise kernel. The new feature of this noise kernel is that it depends on $\Sigma$, which becomes $f(x)$ when the limit $x_{+}-x_{-}\rightarrow 0$ is taken. In other words it depends on the past history of the Brownian particle. Indeed as one can check that the dependence of $N_{2}(s,s';\Sigma)$ on $\Sigma(s'')$ is for $s,s'\geq s''\geq 0$.

Another new feature of the stochastic force we have obtained is that the correlator for three $\xi$'s is nonzero.
\begin{eqnarray}
\left\langle\xi(s)\xi(s')\xi(s'')\right\rangle=\int D\xi\, P[\xi]\,\xi(s)\xi(s')\xi(s'')
=\hbar^{2}  N_{3}(s,s',s'').
\end{eqnarray}
This non-gaussianity of the correlators might be relevant when one applies this higher order consideration to field theory, such as, for example, to the inflaton scalar field in the inflationary de Sitter spacetime. In this case the non-gaussianity of the stochastic force correlators will be manifested in those of the inflaton field operators. This will in turn be related to the non-gaussianity of the cosmic microwave background radiation anisotropy spectrum \cite{ChoNg20,ChoNg26}.

From these correlators one could also study the behaviors of the probability density $P[\xi]$. To the quadratic order of the influence action, or to order $\lambda^{2}$, the stochastic force is gaussian since correlators with odd numbers of $\xi$'s vanish. This implies that to this order $P[\xi]$ is a gaussian functional, that is,
\begin{eqnarray}
P^{(0)}[\xi]=Ce^{-\frac{1}{2\hbar}\int_{0}^{t}ds\int_{0}^{t}ds'\,\xi(s)\left(N_{2}^{(0)}(s,s')\right)^{-1}\xi(s')},
\end{eqnarray}
where $C$ is the normalization constant
\begin{eqnarray}
C=\left(\int D\xi\, e^{-\frac{1}{2\hbar}\int_{0}^{t}ds\int_{0}^{t}ds'\,\xi(s)\left(N_{2}^{(0)}(s,s')\right)^{-1}\xi(s')}\right)^{-1}.
\end{eqnarray}
Here the inverse bifunction $\left(N_{2}^{(0)}(s,s')\right)^{-1}$ is defined by
\begin{eqnarray}
\int_{0}^{t}ds''\, N_{2}^{(0)}(s,s'')\left(N_{2}^{(0)}(s'',s')\right)^{-1}=\delta(s-s').
\end{eqnarray}
From Eqs.~(\ref{nu}) and (\ref{noisen2}), we can express  the noise kernel $N_{2}^{(0)}(s,s')$ in the spectral form.
\begin{eqnarray}
    N_{2}^{(0)}(s,s')=\int_{0}^{\infty}\,d\omega\,I(\omega)\,\cos[2\omega(s-s')]
\end{eqnarray}
where
\begin{eqnarray}
    I(\omega)=\sum_{n}\,\delta(\omega-\omega_{n})\frac{\lambda^{2}\hbar(C_{n1}-C_{n2})^{2}}{2m_{n}^{2}\omega_{n}^{2}}
\end{eqnarray}
It is then possible to express the inverse bifunction in this spectral form as
\begin{eqnarray}
    \left(N_{2}^{(0)}(s'',s')\right)^{-1}=\int_{0}^{\infty}d\omega\left(\frac{4}{\pi^{2}I(\omega)}\right)\cos[2\omega(s-s')]
\end{eqnarray}

To order $\lambda^{3}$ the probability density functional will no longer be gaussian. The exact form of this probability density functional is not easy to find, especially when we have only a few correlators of the stochastic force. Actually the problem of constructing the probability density functional from  various moments has been discussed in detail in \cite{FFR1} for the two-dimensional case and in \cite{FFR2} for the four-dimensional case. Here we shall follow a more straightforward perturbative approach. To order $\lambda^{3}$ we can write schematically
\begin{eqnarray}
P[\xi]=P^{(0)}[\xi]\left(C_{0}+\int C_{1}\xi+\int C_{2}\xi\xi+\int C_{3}\xi\xi\xi\right).
\end{eqnarray}
To the quadratic order, $P[\xi]=P^{(0)}[\xi]$ and we have $C_{0}^{(0)}=1$, $C_{1}^{(0)}=C_{2}^{(0)}=C_{3}^{(0)}=0$. 

To the next order, that is, to order $\lambda^{3}$, we can evaluate $C_{0}^{(1)}$, $C_{1}^{(1)}$, $C_{2}^{(1)}$, and $C_{3}^{(1)}$ by examining the correlators of the stochastic force to this order. In particular, from
$
\left\langle 1\right\rangle^{(1)}=0$,
we obtain
\begin{eqnarray}\label{C0}
C_{0}^{(1)}+\int ds_{1} ds_{2}\, C_{2}^{(1)}(s_{1},s_{2})\left\langle\xi(s_{1})\xi(s_{2})\right\rangle^{(0)}=0.
\end{eqnarray}
From
$
\left\langle \xi(s)\right\rangle^{(1)}=0$,
we have
\begin{eqnarray}\label{C1}
\int ds_{1}\, C_{1}^{(1)}(s_{1})\left\langle\xi(s)\xi(s_{1})\right\rangle^{(0)}
+\int ds_{1}ds_{2}ds_{3}\, C_{3}^{(1)}(s_{1},s_{2},s_{3})\left\langle\xi(s)\xi(s_{1})\xi(s_{2})\xi(s_{3})\right\rangle^{(0)}=0.\nonumber\\
\end{eqnarray}
From
$
\left\langle \xi(s)\xi(s')\right\rangle^{(1)}=\hbar N_{2}^{(1)}(s,s')$,
we have
\begin{eqnarray}\label{C2}
C_{0}^{(1)}\left\langle\xi(s)\xi(s')\right\rangle^{(0)}
+\int ds_{1}ds_{2}\, C_{2}^{(1)}(s_{1},s_{2})\left\langle\xi(s)\xi(s')\xi(s_{1})\xi(s_{2})\right\rangle^{(0)}=\hbar N_{2}^{(1)}(s,s').
\end{eqnarray}
Finally, from
$
\left\langle \xi(s)\xi(s')\xi(s'')\right\rangle^{(1)}=\hbar^{2} N_{3}^{(1)}(s,s',s'')$,
we have
\begin{eqnarray}\label{C3}
&&\int ds_{1}\, C_{1}^{(1)}(s_{1})\left\langle\xi(s)\xi(s')\xi(s'')\xi(s_{1})\right\rangle^{(0)}\nonumber\\
&&\ \ +\int ds_{1}ds_{2}ds_{3}\, C_{3}^{(1)}(s_{1},s_{2},s_{3})\left\langle\xi(s)\xi(s')\xi(s'')\xi(s_{1})\xi(s_{2})\xi(s_{3})\right\rangle^{(0)}
=\hbar^{2} N_{3}^{(1)}(s,s',s'').\nonumber\\
\end{eqnarray}
To solve these equations, we note that to the lowest order the probability density $P^{(0)}[\xi]$ is gaussian. Hence, we have
\begin{eqnarray}
\left\langle\xi(s_{1})\xi(s_{2})\xi(s_{3})\xi(s_{4})\right\rangle^{(0)}&=&\hbar^{2}\big[N_{2}^{(0)}(s_{1},s_{2})N_{2}^{(0)}(s_{3},s_{4})+N_{2}^{(0)}(s_{1},s_{3})N_{2}^{(0)}(s_{2},s_{4})\nonumber\\
&&\hskip 20pt +N_{2}^{(0)}(s_{1},s_{4})N_{2}^{(0)}(s_{2},s_{3})\big]
\end{eqnarray}
where we have made use of Eq.~(\ref{n20}). Similarly, we have
\begin{eqnarray}    \left\langle\xi(s_{1})\xi(s_{2})\xi(s_{3})\xi(s_{4})\xi(s_{5})\xi(s_{6})\right\rangle^{(0)}
&=&\hbar^{3}\big[N_{2}^{(0)}(s_{1},s_{2})N_{2}^{(0)}(s_{3},s_{4})N_{2}^{(0)}(s_{5},s_{6})\nonumber\\
&&\hskip 20pt+{\rm 14\ permutations}\big]
\end{eqnarray}
The 14 permutations include terms like $(12)(35)(46)$, $(12)(36)(45)$, etc.

From these results and the relations in Eqs.~(\ref{C0}) to (\ref{C3}), we can solve for $C_{i}^{(1)}$, $i=0,1,2,3$. Finally, we obtain
\begin{eqnarray}
C_{0}^{(1)}&=&-\frac{1}{2}\int ds_{1}ds_{2}\, N_{2}^{(1)}(s_{1},s_{2};\Sigma)(N_{2}^{(0)}(s_{1},s_{2}))^{-1}\nonumber\\
C_{1}^{(1)}(s)&=&-\frac{1}{2}\int ds_{1}ds_{2}ds_{3}\, N_{3}^{(1)}(s_{1},s_{2},s_{3})(N_{2}^{(0)}(s,s_{1}))^{-1}(N_{2}^{(0)}(s_{2},s_{3}))^{-1}\nonumber\\
C_{2}^{(1)}(s,s')&=&\frac{1}{2\hbar}\int ds_{1}ds_{2}\, N_{2}^{(1)}(s_{1},s_{2};\Sigma)(N_{2}^{(0)}(s_{1},s))^{-1}(N_{2}^{(0)}(s_{2},s'))^{-1}\nonumber\\
C_{3}^{(1)}(s,s',s'')&=&\frac{1}{6\hbar}\int ds_{1}ds_{2}ds_{3}\, N_{3}^{(1)}(s_{1},s_{2},s_{3})(N_{2}^{(0)}(s_{1},s))^{-1}(N_{2}^{(0)}(s_{2},s'))^{-1}(N_{2}^{(0)}(s_{3},s''))^{-1}\nonumber\\
\end{eqnarray}

%
%
%
\section{Modified Fluctuation-Dissipation Relation}\label{sec:FDR}

In Eq.~(\ref{effact}), the term proportional to $\Delta(s)$ is related to the dissipation effect. We can rewrite this term as
\begin{eqnarray}
-\int_{0}^{t}ds\,\Delta(s)\,J(s;\Sigma)=-2\int_{0}^{t}ds\int_{0}^{s}ds'\,\Delta(s)\,\tilde{\gamma}(s,s';\Sigma)\dot{\Sigma}(s'),\label{defdamp}
\end{eqnarray}
where $\tilde{\gamma}(s,s';\Sigma)$ can be regarded as the damping kernel to all orders of $\lambda$. 

To the order $\lambda^{3}$ that we are working with, 
\begin{eqnarray}
\tilde{\gamma}(s,s';\Sigma)=\gamma^{(0)}(s,s')+\gamma^{(1)}(s,s';\Sigma)+\cdots
\end{eqnarray}
where
\begin{eqnarray}
\gamma^{(0)}(s,s')&=&\lambda^{2}\sum_{n}\left(\frac{\hbar}{\omega_{n}}\right)\,(C_{n1}-C_{n2})^{2}\left(\nu_{n}^{2}(s-s')-\mu_{n}^{2}(s-s')\right)=\gamma(s-s'),\nonumber\\
\gamma^{(1)}(s,s';\Sigma)&=&\lambda^{3}\sum_{n}\left(\frac{8\hbar}{\omega_{n}}\right)(C_{n1}-C_{n2})\nonumber\\
&&\ \ \bigg\{-\left(\frac{C_{n2}^{2}}{2m_{n}\omega_{n}^{2}}\right)
\left[\nu_{n}^{2}(s-s')-\mu_{n}^{2}(s-s')\right]\left(\Sigma(s)+\Sigma(s')\right)\nonumber\\
&&\ \ \ \ \ \ \ +(C_{n1}^{2}-C_{n2}^{2})\int_{s'}^{s}ds_{1}\,\mu_{n}(s-s_{1})
\Big[\nu_{n}(s-s')\nu_{n}(s'-s_{1})\nonumber\\
&&\hskip 200pt+\mu_{n}(s-s')\mu_{n}(s'-s_{1})\Big]\Sigma(s_{1})\bigg\}.\nonumber\\
\end{eqnarray}
We have neglected surface terms in deriving $\tilde{\gamma}(s,s';\Sigma)$. The terms at the upper limit is related to renormalization, while the lower limit ones are related to initial kicks without a smooth switching. Discussions on these issues are given in the paper by Fleming, Roura and Hu \cite{FRH}.
$\gamma^{(0)}(s,s')$ here is just the damping kernel in the quadratic order in Eq.~(\ref{diszero}) which is related to the dissipation kernel $\mu(s-s')$ by Eq.~(\ref{mugamma}). While to all orders, the damping kernel $\tilde{\gamma}(s,s';\Sigma)$ depends on $\Sigma$ which will turn into $f(x)$ in the limit $x_{+}-x_{-}\rightarrow 0$. Hence, like the two point correlation of the stochastic force, it also depends on the past history of the Brownian particle.

The damping kernel $\tilde{\gamma}(s,s';\Sigma)$ and the noise kernel $N_{2}(s,s';\Sigma)$ are related by the FDR
\begin{eqnarray}\label{FDR1}
N_{2}(s,s';\Sigma)=\int_{-\infty}^{\infty}ds_{1}\, K(s,s_{1})\,\tilde{\gamma}(s_{1},s';\Sigma).
\end{eqnarray}
where $K(s,s')$ is the dissipation-fluctuation kernel.
To the lowest order in $\lambda$, the FDR is given by
\begin{eqnarray}
N_{2}^{(0)}(s,s')=\int_{-\infty}^{\infty}ds_{1}\, K(s,s_{1})\,\gamma^{(0)}(s_{1},s'),
\end{eqnarray}
Since both 
\begin{eqnarray}
    N_{2}^{(0)},\gamma^{(0)}\sim\lambda^{2}\,\sum_{n}\hbar\,(C_{n1}-C_{n2})^{2}\left[\nu_{n}^{2}(s-s')-\mu_{n}^{2}(s-s')\right]
\end{eqnarray}
one can readily obtain
\begin{eqnarray}\label{KFDR}
K(s,s')=\int_{0}^{\infty}\frac{d\omega}{\pi}\omega\cos[\omega(s-s')].
\end{eqnarray}
which is the same as the one in \cite{HPZ93} valid for zero temperature.

To the next order we can see that $N_{2}^{(1)}(s,s';\Sigma)$ and $\gamma^{(1)}(s,s';\Sigma)$ both have contributions
\begin{eqnarray}
    N_{2}^{(1)},\gamma^{(1)}\sim -\lambda^{3}\sum_{n}\hbar\,(C_{n1}-C_{n2})\left(\frac{C_{n2}^{2}}{m_{n}\omega_{n}^{2}}\right)
\left[\nu_{n}^{2}(s-s')-\mu_{n}^{2}(s-s')\right]\left(\Sigma(s)+\Sigma(s')\right)\nonumber\\
\end{eqnarray}
and 
\begin{eqnarray}
    N_{2}^{(1)},\gamma^{(1)}&\sim& \lambda^{3}\sum_{n}\hbar\,(C_{n1}-C_{n2})(C_{n1}^{2}-C_{n2}^{2})\nonumber\\
    &&\int\ ds_{1}\,\mu_{n}(s-s_{1})
\Big[\nu_{n}(s-s')\nu_{n}(s'-s_{1})+\mu_{n}(s-s')\mu_{n}(s'-s_{1})\Big]\Sigma(s_{1})\nonumber\\
\end{eqnarray}
Hence, we see that the dissipation-fluctuation kernel $K(s,s')$ is still given by Eq.~(\ref{KFDR}) to this order. It is likely that this also holds at higher orders. Accordingly, the FDR between $N_{2}(s,s';\Sigma)$ and $\tilde{\gamma}(s,s';\Sigma)$ is determined by Eq.~(\ref{FDR1}) in conjunction with the kernel in Eq.~(\ref{KFDR}).


%
%

\section{Nonlinear Langevin equation}\label{sec:FDR}

With the consideration on the stochastic force as well as the dissipation effect, the influence action in Eq.~(\ref{effact}) can be rewritten in a more suggestive way as follows.
\begin{eqnarray}
\Gamma[x_{+},x_{-}]&=&\int_{0}^{t}ds\left[\frac{1}{2}M\dot{x}_{+}^{2}-\tilde{V}(x_{+})\right]
-\int_{0}^{t}ds\left[\frac{1}{2}M\dot{x}_{-}^{2}-\tilde{V}(x_{-})\right]\nonumber\\
&&\ \ -2\int_{0}^{t}ds\int_{0}^{s}ds'\, \Delta(s)\tilde{\gamma}(s,s';\Sigma)\dot{\Sigma}(s')
+\int_{0}^{t}ds\, \Delta(s)\xi(s),
\end{eqnarray}
where $\tilde{V}_{1,2}(x)=V(x)+\delta V_{1,2}(x)$ is the renormalized potential. The ellipsis represents terms of higher orders in $\lambda$. 

From this effective action one can derive the corresponding equation of motion for $x$.
\begin{eqnarray}
M\ddot{x}+\tilde{V}_{1}'(x)+\int_{0}^{s}ds'\, \tilde{\gamma}(s,s';f(x))f'(x(s))f'(x(s'))\,\dot{x}(s')=\,f'(x(s))\xi(s).\label{LanEq1}
\end{eqnarray}
This is in the form of a Langevin equation. Since both the damping kernel $\tilde{\gamma}(s,s';f(x))$ and the probability density $P[\xi]$ of the stochastic force $\xi(s)$ depend on the past history of the Brownian particle $x(s)$ with $t>s$, we have a nonlinear Langevin equation.

Next, we would like to examine the solutions to this equation in some details. 
Taking the simple case of $f(x)=x$ and $\tilde{V}_{1}=M\tilde{\Omega}^{2}x^{2}/2$, Eq.~(\ref{LanEq1}) can be written as 
\begin{eqnarray}
M\ddot{x}+M\tilde{\Omega}^{2}x+J(s;x)=\,\xi(s).\label{LanEq2}
\end{eqnarray}
where we have used the functional $J(s;\Sigma)$ instead of $\tilde{\gamma}(s,s';\Sigma)$ as given by the effective action in Eq.~(\ref{effact}). From Eqs.~(\ref{Jkernel}) and (\ref{Jkernels}), the functional $J(s;x)$ can be expanded as
\begin{eqnarray}
J^{(0)}(s;x)&=&\int_{0}^{t}ds'\,J^{(0)}(s,s')x(s'),\nonumber\\
J^{(1)}(s;x)&=&\int_{0}^{t}ds'\int_{0}^{t}ds''\,J^{(1)}(s,s',s'')x(s')x(s''),
\end{eqnarray}
where 
$J^{(0)}(s,s')$ and $J^{(1)}(s,s',s'')$ are defined in Eq.~(\ref{Jkernels}).
Then we write the equation of motion in Eq.~(\ref{LanEq2}) in a perturbative manner as
\begin{eqnarray}
M\ddot{x}+M\tilde{\Omega}^{2}x+\int_{0}^{t}ds'\,J^{(0)}(s,s')x(s')=\xi-\int_{0}^{t}ds'\int_{0}^{t}ds''\,J^{(1)}(s,s',s'')x(s')x(s'').\label{LanEq3}
\end{eqnarray}

To solve this equation of motion perturbatively, we write 
\begin{eqnarray}
    x(s)=x^{(0)}(s)+x^{(1)}(s)+\cdots
\end{eqnarray}
We first consider the homogeneous equation to the lowest order by neglecting the $J^{(1)}$ term.
\begin{eqnarray}
M\ddot{x}^{(0)}+M\tilde{\Omega}^{2}x^{(0)}+\int_{0}^{t}ds'\,J^{(0)}(s,s')x^{(0)}(s')=0.\label{x0Eq}
\end{eqnarray}
Following \cite{HPZ92} we introduce two functions $u_{1}(s)$ and $u_{2}(s)$ which satisfy the above equation with the boundary conditions
\begin{eqnarray}
&&u_{1}(s=0)=1,\ \ u_{1}(s=t)=0,\nonumber\\
&&u_{2}(s=0)=0,\ \ u_{2}(s=t)=1.
\end{eqnarray}
Then the homogeneous solution $x^{(0)}_{h1}(s)$ to Eq.~(\ref{x0Eq}) with the initial condition
\begin{eqnarray}
x^{(0)}_{h1}(s=0)=x_{0},\ \ \dot{x}^{(0)}_{h}(s=0)=\frac{p_{0}}{M},
\end{eqnarray}
is given by
\begin{eqnarray}
x^{(0)}_{h1}(s)=\left[u_{1}(s)-\frac{\dot{u}_{1}(0)}{\dot{u}_{2}(0)}u_{2}(s)\right]x_{0}+\frac{u_{2}(s)}{\dot{u}_{2}(0)}\left(\frac{p_{0}}{M}\right).
\end{eqnarray}
With the stochastic force term $\xi(s)$, the lowest order solution to the equation of motion is just
\begin{eqnarray}
x^{(0)}(s)=x_{h1}^{(0)}(s)+\int_{0}^{t}ds' \,G_{ret}(s,s')\xi(s'),
\end{eqnarray}
where the retarded Green's function $G_{ret}(s,s')$ can be constructed using the functions $u_{1}(s)$ and $u_{2}(s)$,
\begin{eqnarray}
G_{ret}(s,s')=\frac{1}{M}\left[\frac{u_{1}(s)u_{2}(s')-u_{1}(s')u_{2}(s)}{\dot{u}_{1}(s')u_{2}(s')-u_{1}(s')\dot{u}_{2}(s')}\right]\theta(s-s').
\end{eqnarray}

Now we are ready to give the solution $x(s)$ of the equation of motion in Eq.~(\ref{LanEq3}) to the next order in $\lambda$. Using the retarded Green's function we can express
\begin{eqnarray}
x^{(1)}(s)=-\int_{0}^{t}ds'\int_{0}^{t}ds''\int_{0}^{t}ds'''\,G_{ret}(s,s')J^{(1)}(s',s'',s''')\,x^{(0)}(s'')\,x^{(0)}(s''').
\end{eqnarray}
Note that the initial conditions imposed on the solution $x(s)$ we just described are
$x(s=0)=x_{0}$ and $\dot{x}(s=0)=p_{0}/M$.

In a similar fashion we can also find the solution $x(s)$ with a different set of initial conditions,
\begin{equation}
x(s=t)=x_{t},\ \ \dot{x}(s=t)=\frac{p_{t}}{M}.
\end{equation}
In this case 
\begin{eqnarray}
x^{(0)}(s)=x^{(0)}_{h2}(s)+\int_{0}^{t}ds'\,G_{adv}(s,s')\xi(s'),
\end{eqnarray}
where the homogeneous solution
\begin{eqnarray}
x^{(0)}_{h2}(s)=\left[u_{2}(s)-\frac{\dot{u}_{2}(t)}{\dot{u}_{1}(t)}u_{1}(s)\right]x_{t}+\frac{u_{1}(s)}{\dot{u}_{1}(t)}\left(\frac{p_{t}}{M}\right)
\end{eqnarray}
and the advanced Green's function
\begin{eqnarray}
G_{adv}(s,s')=-\frac{1}{M}\left[\frac{u_{1}(s)u_{2}(s')-u_{1}(s')u_{2}(s)}{\dot{u}_{1}(s')u_{2}(s')-u_{1}(s')\dot{u}_{2}(s')}\right]\theta(s'-s).
\end{eqnarray}
To the next order in $\lambda$,
\begin{equation}
x^{(1)}(s)=-\int_{0}^{t}ds'\int_{0}^{t}ds''\int_{0}^{t}ds'''\,G_{adv}(s,s')J^{(1)}(s',s'',s''')\,x^{(0)}(s'')\,x^{(0)}(s''').
\end{equation}
In this form one can see the explicit dependence of the solution $x(s)$ on the values $x_{t}$ and $p_{t}$ at time $t$. This expression will be useful when we derive the Fokker-Planck equation of the Wigner function from the Langevin equation in a follow-up paper \cite{ChoHuQOM}.

%
%


%
 
%

\section{Conclusion and Application to Quantum Optomechanics}\label{sec:conclusion}

Extending upon the foundational research of Hu, Paz, and Zhang \cite{HPZ93}, this study investigates a QBM model featuring a single oscillator system coupling nonlinearly with an environment of harmonic oscillators. The coupling term involves both position and momentum variables as given in Eq.~(\ref{Sint1}). Utilizing the closed-time-path formalism, we derive the influence action through a perturbative expansion of the dimensionless coupling constant $\lambda$. Within this framework, terms in the influence action that are quadratic or higher in $\Delta(s)\equiv f(x_{+}(s))-f(x_{-}(s))$ represent the noise kernel, whereas linear terms define the dissipation kernel. One critical finding is that the noise kernel is non-Gaussian, giving rise to non-vanishing three-point correlation function of the corresponding stochastic force. Such non-Gaussian signatures offer significant utility in quantum field theory, particularly within the context of early universe cosmological models. Furthermore, we have formulated a modified FDR ensuring the consistency of our findings at higher perturbative orders. Finally, the key finding in this paper is the derivation of a nonlinear Langevin equation which will be useful in various applications to open quantum system problems.
Note the Langevin equation presented here is of a semiclassical nature because while the nonGaussian noise is quantum the Brownian particle's motion is classical. Perturbative solutions with different boundary conditions to this equation are provided using retarded and advanced Green’s functions, setting the stage for obtaining the quantum master or Fokker-Planck equations presently under study.

In the introduction we mentioned QOM as an area where methods and results from our present work may be of use. QOM is not only a frontier field of quantum optics involving moving atoms or mirrors, it is also a useful platform for gravitational wave research, such as seeking ways for noise reduction and the detection of the quantum features of gravity.  We add a couple of remarks here, as points for further development.

In relation to actual QOM experimental setups our theoretical studies need be modified in at least two aspects: (i)   \textit{mode-coupling}. Here, the field is a free field, not a cavity field, thus we will not see mode-coupling; (ii) \textit{back-action}. The present set-up of the open system interacting with its environment and the environment acting back on the system dynamics is of the \textit{laissez faire} kind, meaning, the two parties are free to  negotiate between themselves, and only so, as they form a closed system with no external influences.  Without an external agent driving the mirror in motion, any process involving fluctuations of the quantum field (quantum noises) being  parametrically amplified by the motion of a mirror, such as particle creation in the DCE, will be small, and in turn, the back-action effects will be weak. Regardless of the absence or presence of a driving force\footnote{An example of how a time-dependent field applied to both the system particle and the bath affects the modified fluctuation-dissipation relation, see, e.g., \cite{Zaccone}.} and the magnitude of the back-action, the effects of back-action noises can be treated by the formulation presented here because it respects full self-consistency, with proven FDRs providing an effective guardrail of this important, yet often neglected or downplayed, requisite.  

In our next paper using the same model we shall derive the quantum master and  Fokker-Planck equations for the dynamics of a  quantum harmonic oscillator system nonlinearly coupled to an N-oscillator bath. Using these equations for the study of issues in QOM involves putting a quantum field in a cavity and introducing an external agent to set one of the mirrors in motion. In essence this is a study of the DCE of a moving mirror interacting nonlinearly with a quantum field. Because the influence functional method incorporates the back-action of nonGaussian quantum noise in a self-consistent manner, it would be interesting to compare the results with the traditional and popular radiation pressure approach \cite{Law,But25}.

%
\acknowledgments HTC was supported in part by the National Science and Technology Council (NSTC) of Taiwan, Republic of China, under Grant Nos.~NSTC 114-2112-M-032-007. BLH enjoyed the warm hospitality of Profs.~Hsiang-Nan Li and Kin-Wang Ng and colleagues at the Institute of Physics, Academia Sinica when this work began and Prof.~Chong-Sun Chu at the National Tsing Hua University when this work was completed.   

%

\end{document}